\documentclass[12pt]{article}
\usepackage{amsfonts}
\usepackage{epsfig}
\newcommand{\beq}{\begin{eqnarray}}
 \newcommand{\eeq}{\end{eqnarray}}
\newcommand{\be}{\begin{equation}}
 \newcommand{\ee}{\end{equation}}

\def\fun#1#2{\lower3.6pt\vbox{\baselineskip0pt\lineskip.9pt
\ialign{$\mathsurround=0pt#1\hfil ##\hfil$\crcr#2\crcr\sim\crcr}}}

\newcommand{{\SD}}{\rm SD}

\newcommand{\veH}{\mbox{\boldmath${\rm H}$}}
\newcommand{\veE}{\mbox{\boldmath${\rm E}$}}

\newcommand{\lan}{\langle}
\newcommand{\ran}{\rangle}

\begin{document}
\raggedbottom

\title{Nonperturbative QCD: confinement and deconfinement
}

\author{ Yu.A.Simonov\\[5pt]
{\it \large State Research
Center}\\{\it Institute of Theoretical and Experimental Physics,} \\
{\it Moscow, 117218 Russia}}

\date{}

\maketitle

\begin{abstract}

After a short exposition of field correlators in the QCD vacuum
and the recently discovered Casimir scaling phenomenon, the origin
of confinement in QCD is discussed and two possible mechanisms are
suggested, which can be checked by new lattice measurements.
Screening of confinement due  to sea quarks is discussed and
quantitatively explained. Deconfinement is introduced via the
colorelectric field evaporation and the transition temperature
$T_c$ is found numerically in good agreement with lattice
measurements. The $T_c$ dependence on $N_c$ and $n_f$ is also
predicted and agrees with the recent lattice data.

\end{abstract}

\section{Introduction}

The confinement is known to be the most important QCD dynamics at
large distances preserving stability of  matter and existence  of
our world (for a review see \cite{1}). The force of approximately
15 tons between quark  and antiquark in mesons, or between quark
and  the string junction  in proton is known from the  lattice
with excellent accuracy to be constant up to very small distances
\cite{2}. However theoretical understanding of  this phenomenon
still far from complete despite many efforts during the last
decade. It is a purpose of this talk to describe in some detail
what is understood now in the picture of confinement and
deconfinement and where is the front line of our present
knowledge. The main instrument in what follows is the
gauge-invariant Formalism of Field Correlators (FFC) \cite{3,4},
however references to  particular applications  are done at
appropriate places.

As it is understood now the nonperturbative structure of the QCD
vacuum responsible for mass generation and confinement is
adequately described by gauge-invariant field correlators \be
D^{(n)}(1,...n)\equiv \lan tr F_{\mu_1\nu_1}(x_1) \Phi (x_1, x_2)
F_{\mu_2\nu_2} (x_2)... F_{\mu_n\nu_n} (x_n) \Phi(x_n, x_1)
\ran\label{1} \ee where $\Phi(x,y) = P\exp ig \int^x_y A_\mu
dz_\mu$ is the  parallel transporter, and $\lan ... \ran$ means
the vacuum average with the standard  QCD action.

The ensemble of $D^{(n)}, n=2,...\infty,(D^{(1)}\equiv 0)$ or
equivalently, the ensemble of connected correlators $\bar D^{(n)}$
(called cumulants) contains a complete information for the
quark-antiquark dynamics in the limit of large $N_c$ (at finite
$N_c$ also another set of correlators containing more color traces
is necessary).

A recent lattice data \cite{5} for Wilson loops in different
representations of SU(3) color group have confirmed the notion of
Casimir scaling with very high accuracy (around 1\%), i.e. the
fact that static $Q\bar Q$ potentials are proportional to the
quadratic  Casimir operator. As it was argued in \cite{6} it is
the lowest field correlator  $D^{(2)}$ which  has the property of
Casimir scaling, while all higher correlators violate this
property and hence are strongly suppressed in the vacuum. This
remarkable observation makes the picture of the QCD vacuum rather
simple, it can be called the Gaussian Stochastic Vacuum (GSV) and
it also enables one to use the lowest correlator known from
lattice calculations \cite{7} for all nonperturbative dynamical
calculations, including the study of QCD strings in mesons
\cite{8} and baryons \cite{9}, study of spectra of  mesons,
hybrids, glueballs and baryons (for a review and references see
\cite{10,11}).

At the same time one should stress that to make the theoretical
approach of FFC complete, one should calculate FC in the framework
of the same method using the only QCD parameter, (string tension
$\sigma$ or $\Lambda_{QCD}$), as an input, and in addition to
understand the dynamics of Casimir scaling, i.e. suppression of
higher cumulants.

In this direction  only first steps are done \cite{12} with
encouraging results, in particular the smallness of the gluonic
correlation length $T_g$, entering FC, is understood from the
gluelump spectrum \cite{13}.

However, the understanding of the Casimir scaling is not yet
complete. On one hand, the contribution of higher correlators
$D^{(n)}$ to the string tension in general can be estimated as
$(\bar FT^2_g)^{n-1} \bar F$, where $\bar F $ is average field
strength, estimated from the condensate to be $\bar F\sim 0.5 $
GeV$^2$, and here we have not taken into account that cumulants
are connected. If this is included, then the Casimir scaling
violation can be shown to arise due to white exchanges between
"dipoles" $\lan F\Phi F\ran $, while  higher cumulants are
suppressed as $1/N_c^2$ (see last ref. in \cite{6}). Altogether
one gets a suppression factor $\sigma^{(4)}/\sigma^{(2)}\approx
(\bar FT^2_g)^2/N_c^2\sim 1\div 3\%$ which gives an
order-of-magnitude agreement with the violation of Casimir scaling
on the lattice \cite{6}.

 Another set of relevant
questions concerns the nature of confinement, i.e. which field
configurations are responsible for confinement. In the GSV it is
the lowest correlator $D^{(2)}$ which confines, therefore the
question is about the nature of field configurations which
saturate $D^{(2)}$. It is the purpose of this talk to discuss
these points one by one, starting with confinement mechanism in
the next section, and temperature deconfinement in section 3.

\section{Mechanism of confinement}

Since $D^{(2)}$ according to  lattice data \cite{5} ensures some
99\% of confinement, one should look more carefully at its
structure,  \cite{4}, namely $$ \bar
D^{(2)}_{\mu_1\nu_1,\mu_2\nu_2} (x_1, x_2)\equiv
\frac{g^2}{N_c}\lan tr F_{\mu_1\nu_1}(x_1)  \Phi (x_1, x_2)
F_{\mu_2\nu_2} (x_2) \Phi(x_2, x_1)\ran=$$ $$ D(z)
(\delta_{\mu_1\mu_2} \delta_{\nu_1\nu_2} -\delta_{\mu_1\nu_2}
\delta_{\mu_2\nu_1})+\frac{1}{2}[\partial_{\mu_1}(z_{\mu_2}\delta_{\nu_1\nu_2}-z_{\nu_2}\delta_{\nu_1\mu_2})+
$$ \be +(\mu_i\leftrightarrow \nu_i)] D_1(z),~~ z\equiv x_1-x_2,~~
\partial_\mu=\frac{\partial}{\partial z_\mu}.\label{2}\ee

Using  the nonabelian Stokes theorem and  the cluster expansion
(see \cite{1} for details and references) one has for a large
Wilson loop
\be
\lan W(C)\ran =\frac{1}{N_c} \lan tr P \exp ig \int_C A_\mu d
x_\mu\ran = \exp (-\sigma S_{min})\label{3}\ee where
\be\sigma^{(2)}=\frac12 \int D (u) d^2u\label{4a}\ee  and
$S_{min}$ is the minimal area inside the contour $C$. From
(\ref{4a}) it is clear that $D(z)$ plays the role of the order
parameter for confinement (at least for $N_c\to \infty$) and this
is confirmed by lattice calculations of $D(z)$, where $D(z)$
vanishes abruptly above the critical temperature  $T_c$ \cite{7}.
A further analysis can be done applying to both sides of (\ref{2})
the operator $\frac12 e_{\alpha\beta\mu_1\nu_1}
\frac{\partial}{\partial x_\alpha}$ which yields \cite{14}
$$z_\alpha \frac{\partial D(z)}{\partial z^2}
=f_\alpha^{(1)}+f^{(2)}_\alpha,~~ f^{(1)}_\alpha \equiv
\frac{g^2}{24 N_c} [\lan tr D_\gamma \tilde F_{\gamma\beta} (z)
\Phi (z,0) \tilde F_{\alpha \beta} (0) \Phi(0,z) \ran ] ;$$
\be
 f^{(2)}_\alpha \equiv
\frac{g^2}{24 N_c} [ig \int^z_0 dy_\delta \alpha(y)\lan  tr \tilde
F_{\lambda \beta} (z) \Phi (z,y)  F_{\lambda\delta } (y)
\Phi(y,0)\tilde F_{\alpha\beta} (0) \Phi(0,z) \ran-h.c. ]\label{4}
\ee

We start with the Abelian case where one should replace
$D_\gamma\to \partial_\gamma, ~~ \Phi\to 1$ in (\ref{4}) and the
last two terms inside square brackets, coming from the contour
differentiation, are absent:
\be
z_\alpha \frac{\partial D^{abelian}(z)}{\partial z^2}
=f^{(1)}_\alpha=const \lan j^{(mon)}_\beta (z) \tilde
F_{\alpha\beta} (0)\ran\label{5}\ee
 and $j_\beta^{(mon)} \equiv
\partial_\gamma \tilde F_{\gamma\beta} (z)$ is the monopole
current. Hence  in the Abelian  U(1) case confinement  might be
due to Abelian monopoles and this fact agrees with what is known
about the  Abelian mechanism of confinement on the lattice
\cite{1}.

 In the nonabelian case, however, the corresponding Bianchi
 Identities  (BI)
 \be
 D_\mu\tilde F_{\mu\nu} (x) = J_\nu(x),\label{6}\ee
 where $J_\nu(x)\equiv 0$,
 are usually assumed to hold. In the Abelian projection method
 (see \cite{1} for a discussion ) one separates from $F_{\mu\nu}$
 in some special gauge a singular term, the  color-diagonal
 monopole-type configuration which violates both Abelian and
 nonabelian BI. As it is clear from (\ref{4}),
 (\ref{5}) this procedure clearly supplies a source
 $f^{(1)}_{\alpha}$ which makes $D(z)$ nonzero. One should,
 however, make two remarks in this connection. Firstly, when one
 tries to associate the confining configuration  with  some
 classical monopole, there appear no  magnetic monopoles with
finite selfenergy and stable  \cite{15}. Therefore in
 \cite{16} it was concluded that only quantum   confining configuration
 can survive, which however not described analytically.

 Secondly, and this is a more fundamental difficulty, associating
confining configuration with necessary violation of BI, one
usually abandons the
 connection of $F_{\mu\nu}$ with $A_\mu$,
 $F_{\mu\nu}(A)=\partial_\mu A_\nu-\partial_\nu A_\mu-ig [A_\mu,
 A_\nu]$ adding to this expression extra terms (see e.g. in
 \cite{15}). In this way one finds confinement not for the
 original QCD Lagrangian, but for a modified form.

  In what
 follows we shall  insist on  keeping the original
 expression $F_{\mu\nu}(A)$ intact.
In this case the nonabelian BI reduces to the equation
\be
(\partial_\mu\partial_\alpha-\partial_\alpha\partial_\mu) A_\beta
(x) =const~ e_{\alpha\mu\beta\gamma} J_{\gamma} (x)
=0.\label{7}\ee One can see that violation of BI, Eq. (\ref{7}),
requires a very special form of $A_\beta (x)$. One possible form
of $A_\beta(x)$ is given by the path ordered exponents, e.g.
\be
A_\beta (x) =P\exp [\int^x_C dz_\mu \lambda_\mu (z)] b_\beta
(x)+h.c.\label{8}\ee with noncommuting
$\lambda_\mu,\lambda_\nu,\mu\neq\nu$ and the contour $C$, which
belongs to a wide class of nondifferentiable  contours; $x$ is the
end point of the contour, and $P$ orders $\lambda_\mu(z)$ from
say, right to left with $z$ approaching $x$. Such configurations,
if possible, would make nonzero $J_\gamma (x)$ and contribute to
the first term on the r.h.s. of (\ref{4}), thus yielding nonzero
$D(z)$ and string tension. In this way one would find the first
possible source of confinement -- due to the term $f_\alpha^{(1)}$
in (\ref{4}). If this mechanism is proven, it would physically
imply a monopole-like mechanism of confinement. One should stress
that this does not mean real magnetic monopoles or magnetic fluxes
present in the vacuum, since both violate Casimir scaling
\cite{6}, while $D(z)$ supports it. The second source is due to
the term $f^{(2)}_\alpha$ in (\ref{4}), which is always present
irrespectively of the violation of the BI. To make explicit the
meaning of this term one can take the limit $z\to 0$ and obtains
\cite{14}
\be
\frac{dD(x^2)}{dx^2}|_{x^2\to 0}= \frac{g^3}{96 N_c} f^{abc} \lan
F^a_{\alpha\beta} (0) F^b_{\beta\gamma} (0) F^c_{\gamma\alpha}
(0)\ran.
\label{9}\ee

Since the triple correlator can be written as $e_{ijk} f^{abc}$
$\lan E^a_i(0) E^b_j(0) B^c_k(0)\ran$ one can view this second
mechanism as the creation of magnetic fluxes from electric fluxes,
i.e. the electric fluxes contained in the parallel transporter
when shifted in the process of differentiation, serve as a source
of magnetic flux, thus replacing magnetic monopoles.

To distinguish between these two possibilities one could measure
on the lattice the correlator \cite{17}
\be
\Delta_{\nu\beta}(x,y)\equiv \frac{g^2}{N_c}tr \lan D_\mu \tilde
F_{\mu\nu}(x)
 \Phi(x,y) D_\alpha \tilde F_{\alpha\beta} (y)
 \Phi(y,x)\ran.\label{10}\ee

 The nonzero answer for $\Delta_{\nu\beta}$ would mean that the
 violation of BI is indeed the source of confining configurations and writing
 \be
 \Delta_{\alpha\beta} (x)=
 (\partial_\alpha\partial_\beta-\partial^2\delta_{\alpha\beta} )
 D(x) + \Delta^{(2)}_{\alpha\beta}, \label{11a}\ee
 where $\Delta^{(2)}_{\alpha\beta} $ is the contribution of
 $f^{(2)}_{\alpha}$, one can estimate the role of the first
 mechanism as compared to the second. Here $\Delta_{\alpha\beta}$
  is measured via (\ref{10})
 and $D(x)$ is measured directly from (\ref{2}).

 One should note that the given above formulation of  confinement
 mechanisms is fully gauge invariant and does not need gauge
 fixing.

It is worth mentioning that the FFC and the applicability of the
method are not directly related to the question of the mechanism
of confinement, since FFC exploits field correlators as input, but
in the analytic calculation of $D(x)$ and $D_1(x)$ the problem of
confining configurations becomes essential.

\section{Deconfinement}

Discussion of this phenomena can be done in 3 different
directions: 1) when $N_c=3$ and not infinite, the string
connecting static quark  antiquark breaks up at some distance
$R_b$ (typically $R_b\approx 1.4$ fm)  forming two heavy-light
mesons. The same happens for a light $q\bar q$ pair; 2) when the
temperature $T$ exceeds $T_c$ the color electric string between $Q
$ and $\bar Q$ disappears; 3) when baryon density $\rho_B$ exceeds
critical value $\rho_c$, confinement is believed to disappear. We
shall discuss below only points 1) and 2).

1) For finite $N_c$, e.g. $N_c=3$, the effect of sea quarks is
given by the quark determinant $\det (m+\hat D)$ present in the
integral measure of averaging the Wilson loop in (\ref{3}). This
determinant is  producing additional light-quark loops due to the
heat-kernel representation \be \lan \det (m+\hat D) W(C)\ran= \lan
\exp [-\frac12 tr \int^\infty_0 \frac{ds}{s} (Dz)_{xx} e^{-K}
w_\sigma (C_{xx})]W(C)\ran.\label{11} \ee Expanding the first
exponential in (\ref{11}) one obtains corrections to the static
$Q\bar Q$ potential due to the sea-quark loops $w_\sigma (C_{xx})$
where $C_{xx}$ is the closed contour passing through the point
$x$, which is integrated over in (\ref{11}).

The leading correction is proportional to the loop-loop correlator
\be
\chi(C_1, C_2) =\lan W(C_1) W(C_2)\ran - \lan W(C_1)\ran \lan
(W(C_2)\ran,\label{12}\ee which was calculated in \cite{18}, and
the energetically lowest configuration corresponds to the creation
of a hole in the largest quark loop. These holes produce a partial
screening of the $Q\bar Q$ static potential and consequently a
significant decrease of highly excited meson masses. A detailed
analysis of this situation was done in \cite{19} where the form of
the screened potential was found yield a perfect agreement of
predicted light meson masses with experiment (for $L=0,1,2,3$ and
$n_r=0,1,2,3,4$).

Note that this mechanism of the screened confinement is different
from the two-channel model, since in the former sea-quark loops
are virtual and do not cause actual decay. Physically this
difference results in the relative insensitivity of the meson mass
depletion on the channel quantum numbers and on the proximity to
the decay threshold in the screened confinement picture in
contrast to the two-channel approach.

2) {\underline {The temperature deconfinement}

We start with gluodynamics and use background perturbation theory
for $T>0$ \cite{20} which allows to separate in
$A_\mu=B_\mu+a_\mu$ background gluon field $B_\mu$ from valence
gluons $a_\mu$, both with periodic boundary conditions at
$x_4=n\beta=n/T$.

The~ main~ effect~ of~ nonzero~ $T$ ~is~ that the correlators of
electric fields, ~$ g^2\lan E_i(z)E_k(0)\ran= \delta_{ik}
D^E(z)+O(D_1^E)$, and magnetic fields $ g^2\lan H_i(z)H_k(0)\ran=
\delta_{ik} D^H(z)+O(D_1^H)$ are  different, $D^E(z)\neq D^H(z),$
$D_1^E(z) \neq D_1^H(z)$. The same is true for the condensates,
$\lan \veE^2\ran\neq \lan \veH^2\ran$, which define the vacuum
energy density
\be
\varepsilon_0= -\frac{11}{3} N_c\frac{\alpha_s}{16 \pi}\lan
\veE^2+ \veH^2\ran.
 \label{13}\ee

It is clear that the confinement phase  of gluodynamics  with
$\varepsilon =\varepsilon(1)$ consists of glueballs moving in the
confining background $(\veE^2+\veH^2)$, while the deconfinement
phase with  $\varepsilon=\varepsilon(2)$ represents valence gluons
moving in the deconfined vacuum background $(\veH^2)$. It was
understood already in \cite{21} that it is advantageous (for the
minimum of the Free Energy of the Vacuum (FEV)) to keep magnetic
condensate intact at $T>T_c$, while at $T<T_c$ both electric and
magnetic can be nonzero. This implies that $D^{(H)}$ and the
spacial string tension $\sigma_{sp}$ stay constant across phase
transition which was supported by lattice data \cite{7}. One can
define FEV in two phases including quarks as \cite{20} $$
F(1)/V_3= \varepsilon(1) -\frac{\pi}{30} T^4- ({\rm
higher~mesons})$$
\be
F(2)/V_3= \varepsilon(2) - (N^2_c-1) \frac{T^4\pi^2}{45}\Omega_g
-\frac{7\pi^2}{180} N_c  T^4 n_f \Omega_q +O(N_c^0)\label{14} \ee
where $\Omega_q,\Omega_g$ are perimeter contributions to the quark
and gluon loops respectively. Assuming that magnetic part of
$\varepsilon(1)$ does not change for $0\leq T\leq T_c$ and equal
to the electric one, one has $\varepsilon(1) = \varepsilon_0.~~
\varepsilon (2) =\frac12 \varepsilon_0$ and from $F(1)=F(2)$ one
obtains for the transition temperature \cite{20}:
\be
T_c=\left (\frac{|\varepsilon_0|}{\frac{2\pi^2}{45}
(N_c^2-1)\Omega_g+\frac{7\pi^2}{90} N_c n_f
\Omega_q-\frac{\pi}{15} (n_f\geq2)}\right)^{1/4}.\label{15}\ee
Neglecting the influence of magnetic fields on the quark and gluon
motion yields $\Omega_q=\Omega_g=1$,  and one can compute $T_c$
for different $n_f$ taking $\varepsilon_0$ (\ref{13}) from
standard gluon condensate estimate $\frac{\alpha_s}{\pi} \lan
(F_{\mu\nu}^a)^2\ran =0.012$ GeV$^4$, for $n_f=2,3,4$ and 3 times
more for $n_f=0$ \cite{22}. One obtains in this way from
(\ref{15}), $T_c=240,150,141, 134$ MeV for $n_f=0,2,3,4$ which can
be compared to the  lattice data $T_c =270, 172, 154, 131 $ MeV
respectively. One can see a systematic 10-12\% disagreement
 which can be removed taking standard gluonic condensate \cite{22} 1.5 times larger.

Note that (\ref{15}) gives for $T_c$ the constant values for large
$N_c$, and $T_c$ is only weakly dependent on $N_c$.

Moreover, the phase transition is the first order for large $N_c$.
Both facts agree with recent lattice data \cite{24}. Indeed the
analysis of lattice data at $n_f=0$ in \cite{24} yields:
$\frac{T_c}{\sqrt{\sigma}}=0.582 =\frac{0.43}{N_c^2}$, and for
$\sigma=0.18$ GeV$^2$ it gives  $T_c=(0.246+0.02\frac{9}{N^2_c})$
GeV, which agrees well with (\ref{15}) for $\Omega_q=\Omega_g=1$,
where $T_c=0.243$ GeV and for $n_f=0$ it does not depend on $N_c$,
so that (significant) $N_c$ dependence appears only for $n_f>0$.
 The simple picture described above is sometimes called "the
 Vacuum Evaporation Model" (VEM) and is actually the only known
 model successful in all these features. However, the latent heat for
  $N_c=3,~n_f=2$ comes out in VEM  too high, and to improve the
 situation one should take into  account that $\Omega_q,
 \Omega_g\neq 1 $ above $T_c$ \cite{25} and the contribution of
 other than pion mesons in (\ref{14}) below $T_c$, which was done
 in \cite{26}. Both factors make the phase transition more smooth
 and strongly decrease the specific heat.

In conclusion, it is  clear that the present approach yields a
very  consistent picture of both confinement and deconfinement,
which can be derived from the standard QCD Lagrangian using
background perturbation theory and FFC. In this approach inputs
are $D(x), D_1(x)$, which are known from the lattice data \cite{7}
to be $D(x) = D(0) \exp (-|x|/T_g),~~ D_1(x) =D_1(0)\exp
(-|x|/T_g)$ reducing input to 3 numbers,  $D(0), D_1(0)$ and
$T_g$. Since $D_1(0)\ll D(0)$, one actually has $D(0)$ and $T_g$,
or equivalently $\sigma$ and $T_g$, and most hadron spectrum data
are computed through only $\sigma$ with 10\% accuracy.
Nevertheless the method can be considered as logically consistent,
with the statement that  everything including all field
correlators is computed through the only QCD parameter:
$\Lambda_{QCD} $ or $\sigma$. The work in this direction was
partly done in \cite{12}, \cite{13}. There the correlator length
$T_g$ was  expressed through the gluelump masses and $T_g$ was
found  in the range 0.13-0.17 fm.

Moreover the full correlator $D(x), D_1(x)$ can be computed
through the gluelump Green's function \cite{27}, if one assumes
that  the nonabelian BI are not violated. In the opposite case
there appears the problem of identification and study of the
BI-violating configurations, similar to the study of magnetic
monopole ensemble, if they really exist in the QCD vacuum. Indeed
 the Casimir scaling phenomenon
\cite{5} strongly limits the admixture of any coherent
configurations (like classical solutions for monopoles, dyons and
instantons) with size larger than 0.2 fm \cite{6}, which makes the
popular picture of the QCD vacuum with magnetic monopoles or
central fluxes rather unrealistic. Therefore for confinement one
has either the picture of the BI violating configurations with
very small radius or an alternative picture with vanishing BI and
the gluelumps saturating the field correlators.

Present lattice data cannot unfortunately distinguish between two
alternatives since,  as argued above, the popular abelian
projection method or central vortex projection both measure the
quantities proportional to $D(x)$ or its derivatives,  while only
direct measurement of the correlator $\Delta_{\alpha\beta}$
(\ref{10}) can separate two possibilities, provided a clean
lattice analog of $\Delta_{\alpha\beta} $ is formulated.

The author is grateful to D.V.Antonov, M.I.Polikarpov, F.V.Gubarev
and V.I.Shevchenko for discussions.

This work was supported by INTAS grants 00-110 and 00-366 and RFBR
grants 00-02-17836 and 00-15-96736.

\end{document}